\documentclass[,final]{aipproc}
 \layoutstyle{6x9}
\begin{document}

\title{Why PQ?}
\author{R. D. Peccei \footnote{Invited talk given at the Axion 2010 Conference, January 15-17, 2010, Univ. of Florida, Gainsville, Fl.}}{address={Department of Physics and Astronomy, UCLA, Los Angeles, California, 90095}}

\begin{abstract}
I discuss how the solution of the $U(1)_A$ problem of QCD through the existence of the $\theta$-vacuum gave rise to the strong CP problem. After examining various suggested solutions to this problem, I conclude that the only viable solution still is one that involves the existence of a spontaneously broken chiral symmetry:
$U(1)_{PQ}.$
\end{abstract}

\maketitle

\section{Introduction}

In the 1970s the strong interactions had a puzzling problem, which became
 particularly clear with the development of QCD. In the limit of 
     vanishing quark masses, the QCD Lagrangian for N flavors, has a large 
     global symmetry: $ U(N)_V \times U(N)_A$:
\begin{equation}
 q_f \to [e^{i \alpha_aT_a/2}]_{ff`}q_{f`}~~;~~ q_f \to 
     [e^{i\gamma_5\alpha_aT_a/2}]_{ff`}q_{f`}
 \end{equation}
 Because $m_u, m_d << \Lambda_{\rm QCD}$,  for these quarks the $ m_f  \to 0$ 
 limit is sensible. Thus one expects the strong interactions to be
 approximately $U (2)_V \times U (2)_A $ invariant. Indeed, experimentally, 
               one observes that $ U(2)_V = SU(2)_V  \times U(1)_V$ - which 
 corresponds to isospin times baryon number- is a good approximate 
 symmetry of nature, with nearly degenerate nucleon and pion multiplets in the spectrum.

Because dynamically quark condensates form $(<\bar{u}u>=<\bar{d}d>\neq 0)$, for axial symmetries things are different. These quark condensates break $U(2)_A$ down spontaneously and, thus, there are no mixed parity multiplets in the hadron spectrum. However, because of the spontaneous breakdown of $U(2)_A$ one expects the appearance in the spectrum of approximate  Nambu-Goldstone bosons, with nearly vanishing mass  [ $m \to 0$ as $m_u, m_d \to 0$ ]. For $U(2)_A$ one would expect four such bosons $(\vec{\pi}, \eta)$. Although pions are light, $m_{\pi} \sim 0$, one sees no sign of  another light state in the hadronic spectrum, since  $m^2_{\eta}>> m^2_{\pi}$. Weinberg \cite{Weinberg} dubbed this the $U(1)_A$ problem and suggested that, somehow, there was no $U(1)_A $ symmetry in the strong interactions.

It is useful to describe this conundrum in the language of Chiral Perturbation Theory  which reflects the underlying QCD dynamics for the $(\vec{\pi},\eta)$- sector. Defining $\Sigma= \rm{exp}[i(\vec{\pi }.\vec{\tau} +\eta)/F_{\pi}]$, the dynamics  is described by the effective Lagrangian:
\begin{equation}
L_{\rm{eff}} = \frac{F^2_{\pi}}{4}\rm{Tr } \partial_{\mu}\Sigma\partial^{\mu}\Sigma^{\dagger}  + \frac{F^2_{\pi}m^2_{\pi}}{4}\rm{Tr } (\Sigma +\Sigma^{\dagger}) -\frac{M^2_o}{2}\eta^2
 \end{equation}
The first term contains the chiral invariant interactions of the $(\vec{\pi},\eta)$- sector. The second term reflects the breakdown of the $U(2)_A$ symmetry induced by the quark mass terms, with the pion mass $m^2_{\pi} \sim (m_u+m_d)$. Finally, the last term is an ad hoc contribution which produces an explicit breaking of the $U(1)_A$ symmetry beyond that induced by the quark masses.
Provided $M^2_o >> m^2_{\pi}$ this allows for $m^2_{\eta}>> m^2_{\pi}$. What physics gives rise to this last term is the $U(1)_A$ problem.

 \section{The Resolution of the $U(1)_A$ Problem}
 
The resolution of the $U(1)_A$ problem  came through the realization that the QCD vacuum is more complicated than one thought \cite{tH}. This, in effect, makes $U(1)_A$ not a symmetry of QCD, even though it is an apparent symmetry of its Lagrangian in the limit of vanishing quark masses. However, this more complicated vacuum, gives rise to another problem: the strong CP problem. In essence, as we shall see, the question becomes why is CP not very badly broken in QCD?

The $U(1)_A$ symmetry of QCD in some sense is special, since its associated current, in  the symmetry limit of massless quarks, has a divergence which does not vanish because of the chiral anomaly: \cite{ABJ} 
\begin{equation}
\partial_{\mu}J^{\mu}_5=\frac{g^2}{32{\pi}^2}NF^{\mu\nu}_a\tilde{F}_{a\mu\nu}=  NQ
\end{equation}
where N is the number of massless quarks. Because this divergence does not vanish, in effect, although formally QCD is invariant under a $U(1)_A$ transformation  the chiral anomaly affects the action:
\begin{equation}
\delta W= \alpha \int d^4x \partial_{\mu}J^{\mu}_5= \alpha N\frac{g^2}{32{\pi}^2}\int d^4x F^{\mu\nu}_a\tilde{F}_{a\mu\nu}=\alpha N\int d^4x Q
\end{equation}
 However, matters are not that simple because the gluonic pseudoscalar density Q appearing in the above equation is itself a total divergence, as shown by Bardeen: \cite{Bardeen}
\begin{equation}
F^{\mu\nu}_a\tilde{F}_{a\mu\nu}= \partial_{\mu}K^{\mu}
\end {equation}
\noindent where
\begin{equation}
K^{\mu}=\epsilon^{\mu\alpha\beta\gamma}A_{a\alpha}[F_{a\beta\gamma}-\frac{g}{3}f_{abc}A_{b\beta}A_{c\gamma}]
\end{equation}
\noindent This identity makes $\delta W$ a pure surface integral
\begin{equation}
  \delta W= \alpha N\frac{g^2}{32{\pi}^2}\int  d\sigma_{\mu}K^{\mu}
\end{equation}
\noindent  Hence, using the naive boundary condition that the gauge fields vanish at infinity the integral $\int  d\sigma_{\mu}K^{\mu}$ vanishes and, despite the chiral anomaly, $U(1)_A $ appears again to be a symmetry of QCD!

This conclusion, however, is incorrect. What 't Hooft \cite{tH} showed was that the correct boundary condition to use at infinity is that the gauge field $A^{\mu}_a$ should be either zero or a gauge transformation of zero. It turns out that with these boundary conditions there are gauge configurations for which $\int  d\sigma_{\mu}K^{\mu}\neq 0$ and thus,  indeed, $U(1)_A$ is not a symmetry of QCD. One can show \cite{Peccei} that the configurations with  $\int  d\sigma_{\mu}K^{\mu}\neq 0$  contribute to the vacuum to vacuum transition amplitude for QCD, with the final amplitude being a sum of distinct contribution each characterized by a fixed integer number for the index
\begin{equation}
\nu=\frac{g^2}{32{\pi}^2}\int  d\sigma_{\mu}K^{\mu}=\frac{g^2}{32{\pi}^2}\int d^4x F^{\mu\nu}_a\tilde{F}_{a\mu\nu}= \int d^4x Q
\end{equation}
More precisely, different QCD vacua exist, labeled by an arbitrary angle $\theta$. For the $\theta$-vacuum to $\theta$-vacuum transition amplitude the contributions of each $\nu$ sector are added with a weight of $e^{i\nu\theta}$, so that
\begin{equation}
_+<\theta|\theta>_- =\Sigma_{\nu} e^{i\nu\theta}~ _+<\rm{vac}|\rm{vac}>_-|_{\nu~ \rm{fixed}} 
\end{equation}
The complex weights associated with each $\nu$-sector suggests that, in general, one should expect that T, and CP, is violated in QCD. Indeed, recalling the usual path integral representation for $_+<\rm{vac}|\rm{vac>}_-$ the appearance of the phase factor $e^{i\nu\theta}$ in $_+<\theta|\theta>_-$ simply can be absorbed by adding to the QCD Lagrangian an additional $\theta$-dependent term
\begin{equation}
L_{QCD} \to L_{QCD}  + \theta  \frac{g^2}{32{\pi}^2} F^{\mu\nu}_a\tilde{F}_{a\mu\nu}= L_{QCD} + \theta Q
\end{equation}
Obviously the last term above violates T and CP and, as we shall see, gives rise to the strong CP problem.

 These considerations, however, indeed help resolve the $U(1)_A$ problem. 't Hooft's observations tell us that the topological charge density Q, in effect, acts as a dynamical parameter. Thus, when considering the low energy effective theory for the $\vec{\pi}$ and $\eta$ mesons, one should also include  Q as a relevant dynamical parameter. This was done soon after 't Hooft's work by Di Vecchia and  Veneziano \cite{DVV} who augmented the Chiral Lagrangian describing the low energy behavior of QCD by appropriate Q-depended terms.  Their effective Lagrangian, including terms at most quadratic in Q, reads
\begin{equation}
L_{\rm{eff}} = \frac{F^2_{\pi}}{4}\rm{Tr } \partial_{\mu}\Sigma\partial^{\mu}\Sigma^{\dagger}  + \frac{F^2_{\pi}m^2_{\pi}}{4}\rm{Tr } (\Sigma +\Sigma^{\dagger}) +\frac{iQ}{2} \rm{Tr} [\ln \Sigma-\ln \Sigma^{\dagger}] +\frac{Q^2}{F^2_{\pi}M^2_o} +...
\end{equation}
 The 3rd term in $L_{\rm{eff }}$ above is included to take into account of the anomaly in the $U(1)_A$ current, while the 4th term is the lowest order term in a polynomial in Q. Because Q is essentially a background field in Eq. (10) it can be eliminated through its equation of motion and one finds
\begin{equation}
 Q = -\frac{i}{4 [F^2_{\pi} M^2_o]} \rm{Tr} [\ln \Sigma -\ln \Sigma^{\dagger}] = \frac{\eta}{F_{\pi} M^2_o} + ... 
\end{equation}
Thus the last two terms in $L_{\rm{eff}}$ are equivalent to: $ -\frac{M^2_o}{2}\eta^2 $
which serves to provide an additional gluonic mass term for the $\eta$ meson, solving the $U(1)_A$ problem.

This result emerges also directly  in lattice gauge theory simulations of QCD. That indeed there is a gluonic component to the $\eta$ mass can be seen from lattice simulations of  QCD. A recent paper \cite{CM} for example in 2 +1 flavor QCD  obtains masses for the  $\eta$ and $\eta^{\prime}$ mesons in excellent agreement with experiment. Furthermore, these authors show that these masses remains finite in the limit in which the pion mass goes to zero. That is, in the limit where the u and d quarks are massless there are only 3, not 4, Nambu-Goldstone bosons in QCD.

\section{ The Strong CP Problem and its Possible Resolutions}

The resolution of the $U(1)_A$ problem, however, engenders another problem: the strong CP problem. As we saw, effectively, the QCD vacuum structure adds and extra term to the Lagrangian of QCD: $L_{\theta}=\theta Q $. This term violates P and T, but conserves C, and thus can produce a neutron electric dipole moment of order $d_n \sim e(m_q/M_n^2) \theta$. \footnote{The quark mass factor appears since, as we shall see below, the strong CP problem is absent in the limit of  massless quarks} The strong bound on the neutron electric dipole moment $d_n< 2.9 \times 10^{-26}\rm{ecm}$ \cite{edm} requires the angle $\theta$  to be very small, with best estimates \cite{estimates} giving the bound $\theta< 10^{-9}- 10^{-10}.$ Why the phase angle $\theta$ should be this small is the strong CP problem.

The problem is actually worse if one considers the effect of chiral transformations on the $\theta$-vacuum. One can show that a chiral $U(1)$ transformation, because of the anomaly, in fact changes the $\theta$-vacuum: \cite{JR}
\begin{equation}
e^{i\alpha Q_5} | \theta > = | \theta + \alpha >  
\end{equation}
\noindent If besides QCD one includes the weak interactions, in general the quark mass matrix M is non-diagonal and complex. In a physical basis, one must diagonalize M and in so doing one must, among other things, perform a chiral $U(1)$ transformation  by an angle of Arg det M. As a result of Eq. (13), this changes $\theta$ into
\begin{equation}
\theta_{\rm{total}}= \theta + \rm{Arg~ det M}
\end{equation}
\noindent Thus, in full generality, the strong CP problem can be stated as follows: why is the angle $\theta_{\rm{total}}$, coming from the strong and the weak interactions, so small?

There are three possible "solutions" to the strong CP problem:

\noindent i. Anthropically $\theta_{\rm{total}}$ is small

\noindent ii. CP is spontaneously broken and the induced $\theta_{\rm{total}}$ is small

\noindent iii. A chiral symmetry drives $\theta_{\rm{total}} \to 0$

\noindent In my opinion, only iii. is a viable solution and it necessitates introducing in the Standard Model a new global, spontaneously broken, symmetry:$ U(1)_{PQ}$. \cite{PQ}

\subsection{\it{ i. Anthropic solution}}

It is, of course, possible that, as a result of some anthropic reasons $\theta_{\rm{total}}$ just turns out to be of $O(10^{-10})$. There are, after all, such small ratios in the Standard Model [e.g. $m_e/m_t \sim10^{-6}$]. In my view, what make this explanation unlikely is that the physics of the QCD vacuum (and hence $\theta$) and that of the quark mass matrix ($\rm{Arg~ det M}$) seem totally unrelated. So, why should their sum $\theta_{\rm{total}}$ be a small phase angle?

\subsection{\it{ii. Spontaneously broken CP solution}}

This second possibility is more interesting. In fact, if CP were a symmetry of nature, one can set $\theta=0$ at the Lagrangian level. However, since CP is violated experimentally, this symmetry must be spontaneously broken. This means that an effective angle $\theta$ gets induced back at the loop-level. \cite{spont} But to get $\theta < 10^{-9} -10^{-10}$ one needs, in general, also to insure that $\theta_{1-\rm{loop}}=0$. This is not easy. Furthermore, there are other problems associated with spontaneously broken CP violation which makes this explanation of the strong CP problem unlikely.

Theories with spontaneously broken CP need complex Higgs VEVs and, as Zeldovich, Kobzarev and Okun \cite{ZKO} pointed out, these VEVs give rise to different CP domains in the Universe which are separated by walls with substantial energy density. Because the domain walls are 2-dimensional they dissipate slowly as the Universe cools [$\rho_{\rm{ wall}} \sim T$]. Indeed, if these domains existed, the energy density in the walls would badly overclose the Universe now. Thus, to countenance spontaneous CP violation, the energy scale where CP breaks down has to be greater than the temperature associated with inflation in the Universe [$T_{\rm{inflation}} \sim 10^{10}$ GeV].

Theories where CP is violated at high scales and which have $\theta_{1-\rm{loop}}=0$ exist, \cite{NB} but they are recondite and difficult to reconcile with experiment.  For example, typically the CP violating phases generated at some high scale M induce small phases at low energy, \cite{MMP} reduced by factors of $M_Z/M$. However, this is not what one sees experimentally. All experimental data is in excellent agreement with  the CKM Model where CP is explicitly, not spontaneously, broken and where the CP violating phases are of O(1).

\subsection{\it{iii. A chiral symmetry drives} $\theta_{\rm{total}} \to 0$}

This is a very natural solution to the strong CP problem since, as we saw earlier, a chiral symmetry effectively allows one to rotate one $\theta$-vacua into another. Two possibilities are open in this case. Either one has a real chiral symmetry in the theory, as would happen if the mass of the u-quark vanished [$m_u=0$] as suggested by Kaplan and  Manohar. \cite{KM} Then  all $\theta$-vacuua are equivalent and thus there is no CP violation. Or, as Helen Quinn and I suggested, \cite{PQ} the Standard Model has an additional global $U(1)$ chiral symmetry which is spontaneously broken. In this case one can show that, dynamically, still $\theta_{\rm{total}} \to 0$. I want to argue here that only the PQ solution is tenable.

I always found it difficult to understand why the determinant of the quark mass matrix should vanish, since I could not trace the origin of this chiral symmetry in the quark sector to any physical requirement (except that it should solve the strong CP problem!). However, the " solution"  of the strong CP problem obtained by presuming that $m_u = 0$ is also disfavored by more solid arguments. The original current algebra analysis was done by Leutwyler \cite{L} and he has revisited the subject recently. \cite{L2} The key observation is that the ratio of quark masses is computable in Chiral Perturbation Theory and at leading order, correcting for electromagnetic effects, one arrives at the famous Weinberg formula: \cite{W2}
\begin{equation}
\frac{m_u}{m_d}=\frac{M^2_{K^+} - M^2_{K^0} + 2M^2_{\pi^0}- M^2_{\pi^+}}{ M^2_{K^0} - M^2_{K^+} + M^2_{\pi^+}}= 0.56
\end{equation}
It turns out that this formula does not get large corrections, so the "solution" $m_u = 0$ is really not realized in nature. In his recent paper Leutwyler \cite{L2} has compiled the theoretical predictions for the ratio $m_u/m_d$,  including those coming from lattice QCD simulations. These results make it clear  that the $m_u=0$ "solution" is not supported by data. Indeed, the latest lattice QCD results from the MILC Collaboration rules out $m_u=0$ at $10 \sigma$! \cite{MILC}

Introducing a global $U(1)_{PQ}$ symmetry, which is necessarily spontaneously broken, in effect serves to replaces the static CP violating phase $\theta_{\rm{total}}$ by a dynamical CP conserving axion field $a(x)/f_a$. The axion, as first discussed by Weinberg and Wilczek, \cite{WW} is the Goldstone boson of the broken $U(1)_{PQ}$ symmetry and $f_a$ is the scale of the breaking. It follows then that under $U(1)_{PQ}$ transformations the axion fields just shifts proportional to $f_a$. As a result, if there is indeed a spontaneously broken $U(1)_{PQ}$ symmetry, the low energy Lagrangian of the Standard Model is augmented by interactions of the axion field with the fields in the Standard Model. In particular, to reflect the anomalous nature of the chiral $U(1)_{PQ}$ current there is an interaction of the axion field with the gluonic pseudoscalar density:
\begin{equation}
L_{\rm{anomaly}}= \xi \frac{a}{f_a}  \frac{g^2}{32{\pi}^2} F^{\mu\nu}_a\tilde{F}_{a\mu\nu}= \xi \frac{a}{f_a} Q
\end{equation}
Here $\xi$ is a model dependent parameter reflecting the anomaly of the $U(1)_{PQ}$ current:
\begin{equation}
\partial_{\mu} J^{\mu}_{PQ}= \xi \frac{g^2}{32{\pi}^2} F^{\mu\nu}_a\tilde{F}_{a\mu\nu}
\end{equation}

The Lagrangian term in Eq. (16) acts as an effective potential for the axion field. Effectively, the above term generates an extra contribution to the $\theta$-vacuum angle of  $\xi <a>/f_a$. Because the potential for the axion field is periodic in the total effective vacuum angle $ \theta_{\rm{total}} + \xi<a>/f_a$ minimizing this potential with respect to <a> gives the PQ solution in which the total effective vacuum angle vanishes. Hence the theory written in terms of  a physical axion field $   a_{\rm{phys}}= a- <a>$ has no longer a CP violating $\theta$-term. What remains in the theory is an effective CP conserving interaction of the physical axion field with the gluonic pseudoscalar density Q. If a PQ  symmetry exists the strong CP problem is dynamically resolved, with the CP violating interaction $L_{\theta}=\theta_{\rm{total}}Q$ being replaced by a CP conserving interaction between the axion field and the pseudoscalar density Q: $L_{anomaly}=\xi( a_{\rm{phys}}/f_a)Q$. More than 30 years since it was first suggested, this still appears to be the only truly viable solution to the strong CP problem. 

 \section{Acknowledgments}

It is pleasure to have given this talk at a Symposium in honor of Pierre Sikivie's 60th birthday. Pierre more than anyone I know has kept the axion flame alive and has contributed profound ideas that one day, hopefully, will lead to their detection.

\bibliographystyle{aipproc}

\end{document}